\documentstyle[12pt]{article}

\def\dspace{\baselineskip = .30in}

\begin{document}

\title{An SO(10) Solution to the Puzzle of Quark and Lepton Masses
\footnote{Supported
in part by Department of Energy Grant \#DE-FG02-91ER406267}}

\author{{\bf K.S.Babu and S.M.Barr}\\
Bartol Research Institute\\ University of Delaware\\
Newark, DE 19716}

\date{BA-95-11, hep-ph/9503215}
\maketitle

\begin{abstract}
It is shown that almost all features of the quark and lepton masses
can be satisfactorily and simply explained without family symmetry,
including the threefold mass hierarchy among the generations, and
the relations $m_{\tau}^0 \cong m_b^0$, $m_{\mu}^0 \cong 3 m_s^0$,
$m_e^0 \cong \frac{1}{3} m_d^0$, $m_u^0/m_t^0 \ll m_d^0/m_b^0$,
$\tan \theta_c \cong \surd \overline{m_d^0/m_s^0}$, $V_{cb} \ll
\surd \overline{m_s^0/m_b^0}$, and $V_{ub} \sim V_{cb}V_{us}$.
Various aspects of the group theory of $SO(10)$ play an essential
role in explaining these relations. The form of the mass matrices,
rather than being imposed arbitrarily, emerges naturally from a simple
structure at the unification scale. This structure involves only
vector, spinor and adjoint representations.
There are distinctive and testable predictions for $\tan \beta$ and the
neutrino mixing angles.
\dspace

\end{abstract}
\newpage

\dspace

In this Letter we show that the group theory of $SO(10)$ can provide
a satisfactory explanation of almost all of the features of the quark and
lepton masses and mixing angles$^1$ without family symmetry.  We propose
a model, in two versions, each of which has several testable consequences
for neutrino mixing angles and for the parameter $\tan \beta$.

We propose that the pattern of fermion masses is determined by the
following Yukawa terms in the superpotential$^2$.
\begin{equation}
\begin{array}{ccl}
W & = & M({\bf \overline{16}}\; {\bf 16}) + \sum_{i=1}^3 b_i
({\bf 16}_i {\bf \overline{16}}){\bf 45}_H + \sum_{i=1}^3 a_i ({\bf 16}_i
{\bf 16}) {\bf 10}_H \\
& & \\
& + & d({\bf 10}\; {\bf 10}^{\prime}){\bf 45}_H^{(X)}
+ \sum_{i=1}
^3 c_i ({\bf 16}_i {\bf 10}) {\bf 16}_H + \sum_{i=1}^3 c_i^{\prime}
({\bf 16}_i {\bf 10}^{\prime}) {\bf 16}_H. \\
\end{array}
\end{equation}

\noindent
In addition to the three ordinary families of quarks and leptons,
which are contained in the spinors denoted ${\bf 16}_i$, there is a
pair of spinor and antispinor (${\bf 16}, {\bf \overline{16}}$), and a
pair of vectors (${\bf 10},{\bf 10}^{\prime}$). The Higgs supermultiplets
are distinguished by the subscript `$H$'. We will henceforth call the first
three terms of eq.(1) $W_{spinor}$ and the last three terms $W_{vector}$.

It will be shown that this simple structure explains the following
nine well-known features of the quark and lepton spectrum.

\noindent
{\it The hierarchy:}

\noindent
{\bf (I)} The first generation is much lighter
than the second and third.

\noindent
{\bf (II)} The second generation is much
lighter than the third.

\noindent
{\it The mass ratios:}

\noindent
{\bf (III)} $m_b^0 \cong
m_{\tau}^0$ (see Ref. 3). [The superscript $^0$ refers
throughout to quantities at the unification scale.]

\noindent
{\bf (IV)} $|m_{\mu}^0/m_s^0| \cong 3$.

\noindent
{\bf (V)}
$|m_e^0/m_d^0| \cong |m_{\mu}^0/m_s^0|^{-1} \cong \frac{1}{3}$. (The
Georgi-Jarlskog relations.$^4$)

\noindent
{\bf (VI)} $m_u^0/m_t^0 \ll m_d^0/m_b^0$ and $m_e^0/m_{\tau}^0$.
($10^{-5}$ {\it
versus} respectively $10^{-3}$ and $0.3 \times 10^{-3}$.)

\noindent
{\it The mixing angles:}

\noindent
{\bf (VII)}
$V_{cb}$ is small compared to $\surd \overline{m_s^0/m_b^0}$.

\noindent
{\bf (VIII)}
$\tan \theta_c \cong \surd \overline{m_d^0/m_s^0}$ (see Ref. 5).

\noindent
{\bf (IX)}
$V_{ub} \sim V_{cb}V_{us}$.

In understanding the structure of eq.(1) a crucial point is
that there is no Yukawa term $\sum_{i,j=1}^3 f_{ij}
({\bf 16}_i {\bf 16}_j) {\bf 10}_H$ coupling the ordinary families
to each other$^6$, for the coefficient of such a term, $f_{ij}$, would be
a matrix in `family space', and one would naturally expect the three
eigenvalues of such a matrix to be all of the same order rather than
to exhibit a hierarchical pattern as observed in nature. Instead,
what happens in this model is that the families (${\bf 16}_i$)
couple to each of the `extra' fermion multiplets (${\bf 16},
{\bf \overline{16}},
{\bf 10}, {\bf 10}^{\prime}$) with Yukawa couplings that are
{\it vectors} in family space ($a_i$, $b_i$, $c_i$, $c_i^{\prime}$).
This leads, to first approximation, as will be seen, to a `factorized'
form$^7$ for the mass matrices, $M_{ij} \sim a_i b_j$, which because
it has rank less than three does produce a hierarchy.

Another role played by the `extra' fermions is to allow the ${\bf 45}_H$,
which breaks $SO(10)$, to couple directly to the quarks and leptons
as it could not renormalizably do were there only the ordinary families,
${\bf 16}_i$. This is important because some of the predictions of
minimal $SO(10)$, such as $m_{\mu}^0 = m_s^0$ and $m_c^0/m_t^0 =
m_s^0/m_b^0$, are badly broken in nature.

We will also discuss a variant of this model which is obtained by adding
to the superpotential of eq.(1) the following piece.
\begin{equation}
W_{adjoint} = f({\bf 45}\; {\bf 45}^{\prime}){\bf 45}_H
+ \sum_{i=1}^3 e_i({\bf 16}_i {\bf 45}) {\bf \overline{16}}_H
+ \sum_{i=1}^3 e_i^{\prime}({\bf 16}_i {\bf 45}^{\prime})
{\bf \overline{16}}_H.
\end{equation}

\noindent
The structure of this set of terms is quite analogous to $W_{spinor}$
and $W_{vector}$. Here the `extra' fermions are adjoints (${\bf 45}$,
${\bf 45}^{\prime}$) which couple, as before, to the ordinary families
with coefficients which are vectors in family space ($e_i$,$e_i^{\prime}$).
Note that because the Higgs fields ${\bf 45}_H^{(X)}$ and ${\bf 45}_H$
are antisymmetric tensors, the
${\bf 10}$ and ${\bf 10}^{\prime}$ in $W_{vector}$ must be distinct,
as must the ${\bf 45}$ and ${\bf 45}^{\prime}$ in $W_{adjoint}$.

We will denote the mass matrices of the quarks and leptons by $U$, $D$,
$L$, $N$, and $M_R$. More precisely,
$W_{mass} = \sum_{i,j=1}^3 [u_{Li}^c U_{ij} u_{Lj} + d_{Li}^c D_{ij}
d_{Lj} + l_{Li}^+ L_{ij} l_{Lj}^- + \nu_{Li}^c N_{ij} \nu_{Lj}
+ \nu_{Li}^c (M_R)_{ij} \nu_{Lj}^c].$
The dominant contributions to these matrices are assumed to come from
$W_{spinor}$ and have the exact form$^8$ (up to terms of order
$M_W/M_{GUT} \sim 10^{-14}$)
\begin{equation}
U_0 = aTv \left( \begin{array}{ccc}
0 & 0 & 0 \\
0 & 0 & Q_u \sin \theta / N_u \\
0 & Q_{u^c} \sin \theta / N_{u^c} & (Q_{u^c} + Q_u) \cos \theta /
N_{u^c} N_u
\end{array} \right),
\end{equation}

\begin{equation}
D_0 = aTv' \left( \begin{array}{ccc}
0 & 0 & 0 \\
0 & 0 & Q_d \sin \theta / N_d \\
0 & Q_{d^c} \sin \theta / N_{d^c} & (Q_{d^c} + Q_d) \cos \theta /
N_{d^c} N_d
\end{array} \right),
\end{equation}

\begin{equation}
L_0 = aTv' \left( \begin{array}{ccc}
0 & 0 & 0 \\
0 & 0 & Q_{l^-} \sin \theta / N_{l^-} \\
0 & Q_{l^+} \sin \theta / N_{l^+} & (Q_{l^+} + Q_{l^-}) \cos \theta /
N_{l^+} N_{l^-}
\end{array} \right),
\end{equation}

\begin{equation}
N_0 = aTv \left( \begin{array}{ccc}
0 & 0 & 0 \\
0 & 0 & Q_{\nu} \sin \theta/ N_{\nu} \\
0 & Q_{\nu^c} \sin \theta / N_{\nu^c} & (Q_{\nu^c} + Q_{\nu}) \cos
\theta / N_{\nu^c} N_{\nu}
\end{array} \right).
\end{equation}

\noindent
Here $\theta$ is the angle between the Yukawa coupling constant
vectors $a_i$ and $b_i$ appearing in eq.(1), and $a$ and $b$
are their lengths. $Q$ is an $SO(10)$ generator$^9$ giving the
direction in group space of the vacuum expectation value (VEV) of the
adjoint Higgs field, ${\bf 45}_H$,
\begin{equation}
\langle {\bf 45}_H \rangle = \Omega Q,
\end{equation}
\noindent
and $Q_f$ is the $Q$ charge of the fermion $f$. $\Omega$, like
$M$, is of order $M_{GUT}$, so that we define a dimensionless
ratio $b \Omega/M \equiv T$ which is of order unity. The
$N_f$ are defined by $N_f \equiv
\surd \overline{1 + T^2 \left| Q_f \right| ^2}$. (These factors
of $N_f$ do not play a significant role in what follows. For
$T<1$ they are close to one. We keep them for the sake of exactness.)
Finally, $v$ and $v'$ are the usual $SU(2)_L \times U(1)_Y$-breaking
VEVs of the ${\bf 5}({\bf 10}_H)$ and
$\overline{{\bf 5}}({\bf 10}_H)$ respectively. (Throughout
`${\bf p}({\bf q})$' denotes a ${\bf p}$ of $SU(5)$ contained
in a ${\bf q}$ of $SO(10)$.)

The exact forms given in eqs.(3) -- (6) can be simply derived
by straightforward algebra, but can be more easily understood from
Fig. 1.

\begin{picture}(360,216)
\thicklines
\put(36,144){\vector(1,0){36}}
\put(72,144){\line(1,0){72}}
\put(180,144){\vector(-1,0){36}}
\put(180,144){\vector(1,0){36}}
\put(216,144){\line(1,0){72}}
\put(324,144){\vector(-1,0){36}}
\put(108,96){\line(0,1){12}}
\put(108,114){\vector(0,1){12}}
\put(108,132){\line(0,1){12}}
\put(252,96){\line(0,1){12}}
\put(252,114){\vector(0,1){12}}
\put(252,132){\line(0,1){12}}
\put(175.5,142){$\times$}
\put(103.5,94){$\times$}
\put(247.5,94){$\times$}
\put(94.5,78){$\langle {\bf 10}_H \rangle$}
\put(238.5,78){$\langle {\bf 45}_H \rangle$}
\put(63,153){${\bf 16}_i$}
\put(103.5,153){$a_i$}
\put(139.5,153){${\bf 16}$}
\put(175.5,153){$M$}
\put(207,153){$\overline{{\bf 16}}$}
\put(247.5,153){$b_j$}
\put(279,153){${\bf 16}_j$}
\put(162,42){{\bf Fig. 1}}
\end{picture}

\noindent
By inspection, that diagram gives the expression
\begin{equation}
W_{mass} = \sum_{i,j=1}^3 a_i b_j \frac{\langle {\bf 10}_H \rangle
\langle {\bf 45}_H \rangle }
{M} ({\bf 16}_i {\bf 16}_j );
\end{equation}

\noindent
or $W_{mass} = aT\langle {\bf 10}_H \rangle ( \sum_{i=1}^3
\hat{a}_i {\bf 16}_i)
(\sum_{j=1}^3 \hat{b}_j Q_{16_j} {\bf 16}_j)$, where
$\hat{a}_i \equiv a_i/a$, $\hat{b}_i \equiv b_i/b$. For the up quarks
this gives
\begin{equation}
W^{up}_{mass} = aTv \sum_{i,j=1}^3 [ \hat{a}_i \hat{b}_j Q_u +
\hat{a}_j \hat{b}_i Q_{u^c} ] u_i^c \; u_j,
\end{equation}

\noindent
with similar expressions for the down quarks and leptons. Without
any loss of generality one may choose the axes in family space
so that
\begin{equation}
\begin{array}{ccl}
\hat{a}_i & = & (0, \; \sin \theta, \; \cos \theta), \\
\hat{b}_i & = & (0, \; 0, \; 1),
\end{array}
\end{equation}

\noindent
which when substituted into eq.(9) and its analogues gives the
matrices displayed in eqs.(3) -- (6) with $N_f = 1$. (To build up the
factors $N_f^{-1}$ one must sum over all tree graphs with arbitrary
numbers of superheavy mass or VEV insertions. More simply one can just
do the algebra of integrating out the ${\bf 16}$ and ${\bf
\overline{16}}$.)

Because $\langle {\bf 45}_H\rangle$ cannot break the Standard Model
gauge group,
$Q$ must be a linear combination of the hypercharge and the $SU(5)$-
singlet generator, $X$. ($SO(10) \supset SU(5) \times U(1)_X$.)
For convenience $Q$ will be normalized so that
\begin{equation}
\begin{array}{ccl}
Q & = & (\frac{-1}{5}) X + (\frac{6}{5} z) {Y \over 2} =
2 \; I_{3R} + (\frac{6}{5} \epsilon) \; {Y \over 2},
\end{array}
\end{equation}
\noindent
where $X$ is normalized conventionally, so that $X_{{\bf 10}({\bf 16})}
= 1$, $X_{{\bf \overline{5}}({\bf 16})} = -3$, $X_{{\bf 1}({\bf 16})} = 5$;
and $\epsilon \equiv z - 1$.
Thus, for the fermions contained in the ${\bf 16}$ of $SO(10)$,
\begin{equation}
\begin{array}{ccccl}
Q_u & = & Q_d & = & \frac{1}{5} \epsilon \\
& & Q_{u^c} & = & -1 -\frac{4}{5} \epsilon \\
& & Q_{d^c} & = & 1 + \frac{2}{5} \epsilon \\
Q_{l^-} & = & Q_{\nu} & = & - \frac{3}{5}
\epsilon \\
& & Q_{l^+} & = & 1 + \frac{6}{5} \epsilon \\
& & Q_{\nu^c} & = & -1. \\
\end{array}
\end{equation}

\noindent
We shall assume, for reasons that will become apparent shortly,
that $Q$ is oriented approximately in the $I_{3R}$ direction;
that is, that $|\epsilon| \ll 1$. Then it is seen from their
definitions that the $N_f$ for all the left-handed fermions
($u$, $d$, $l^-$, $\nu$) are very close to one, while for all
the left-handed anti-fermions ($u^c$, $d^c$, $l^+$, $\nu^c$)
 $N_f \cong \surd \overline{1+ T^2} \equiv N$.

Several striking features$^8$ of the mass matrices
in eqs.(3) -- (6) are evident upon inspection.

As a consequence of factorization (cf. eq.(9)) the mass matrices
are rank 2. This explains {\bf (I)}, the lightness of the first
generation. Also explained is {\bf (III)}, $m_b^0 \cong m_{\tau}^0$.
The origin of this relation is the fact that $D_{33} \cong aTv'(Q_{d^c} +
Q_d)\cos \theta/N = aTv'(Q_{l^+} + Q_{l^-})\cos \theta /N
\cong L_{33}$.
The equality of these expressions is no accident, but a consequence of
the fact that $l^+ \; l^-$ and $d^c \; d$ both couple to the same
Higgs doublet, $\phi'$, and thus must have equal charges. On the other hand,
eqs.(4) and (5) show that $|m_{\mu}^0/m_s^0| \neq 1$ but rather
\begin{equation}
\left| \frac{m_{\mu}^0}{m_s^0}\right| \cong \left| \frac{m_{\mu}^0
m_{\tau}^0}
{m_s^0 m_b^0}\right| = \left| \frac{\mbox{det}_{23}L}{\mbox{det}_{23}D}
\right| \cong \left| \frac{Q_{l^+}Q_{l^-}}{Q_{d^c}Q_d}
\right| =
3 \left| \frac{1+ 6 \epsilon/5}{1 + 2 \epsilon/5} \right|.
\end{equation}

\noindent
Thus the empirical relation {\bf (IV)}, $m_{\mu}^0/m_s^0 \cong
3$ is seen to
follow if $Q$ is approximately in the $I_{3R}$ direction; that is,
if $\left| \epsilon \right| \ll 1$. (Cf. eq.(11).)

The $I_{3R}$ direction is
a natural one for the vacuum expectation value of a Higgs field in
the adjoint representation of $SO(10)$. For example,$^{10}$ the superpotential
$-\mu tr(A^2) + \lambda tr(A^4)$, where $A$ is a ${\bf 45}$, only has
$SU(3) \times SU(2) \times U(1)$-invariant solutions in the $I_{3R}$,
$B-L$, and $X$ directions. Other interesting superpotentials$^{11}$
give the same possibilities.
As evidence that adjoint Higgs might well have potentials of this type, it is
significant that the Dimopoulos-Wilczek mechanism$^{12}$, which is
necessary to solve the `doublet-triplet splitting problem' in $SO(10)$
requires the existence of an adjoint Higgs whose vacuum expectation
value is in the $B-L$ direction.

The assumption that $Q$ is approximately in the $I_{3R}$ direction
would also provide a group-theoretical
explanation for two other facts, {\bf (II)} and {\bf (VII)}.
{\bf (II)}, that the second generation is much lighter than the third,
can be seen from eqs.(3) -- (6), which reveal that the (2,3)
elements of all the matrices vanish as $\epsilon \rightarrow 0$,
causing the matrices to become rank 1. For example, $m_s^0/m_b^0
\cong \frac{N}{5} \epsilon \sin^2 \theta$.
That {\bf (VII)}, $V^0_{cb}$ is small compared to
$\surd \overline{m_s^0/m_b^0}$, follows from
$V^0_{cb} \cong {\rm tan}^{-1}(U_{32}/U_{33}) -
{\rm tan}^{-1}(D_{32}/D_{33}) \cong \frac{2}{5} \epsilon
\sin \theta \cos \theta$. In the $\epsilon \rightarrow 0$ limit both
$V_{cb}^0$
and $V_{cb}^0/\surd \overline{m_s^0/m_b^0} \cong \surd \overline{4 \epsilon
/5N} \cos \theta$ vanish. It is well-known that Fritzsch-type
relations$^{13}$
for $V^0_{cb}$ such as $|V^0_{cb}| = |\surd \overline{m_s^0/m_b^0} -
e^{i \alpha}
\surd \overline{m_c^0/m_t^0}|$ fail even though the analogous relation for
the Cabibbo angle works well, because the observed $V^0_{cb}$ is about
a factor of three
too small. Here that smallness is explained as due to the smallness of
$\epsilon$.

There is one potentially troubling feature of the forms given in
eqs.(3) -- (6) and that is that in the small $\epsilon$ limit
the bad `proportionality' prediction of naive $SO(10)$,
namely $m_c^0/m_t^0 \cong m_s^0/m_b^0$, holds. The violation of
this relation will be
explained in an interesting way below.

Taking into account the heretofore neglected terms $W_{vector}$ one
can show that the exact mass matrices (up to terms of order $M_W/M_{GUT}
\sim 10^{-14}$)
coming from eq.(1) are (using eq.(12))
\begin{equation}
U = U_0 = aTv \left( \begin{array}{ccc}
0 & 0 & 0 \\
0 & 0 & \frac{1}{5}\epsilon \sin \theta/N_u \\
0 & -(1+ \frac{4}{5} \epsilon) \sin \theta/N_{u^c}
& -(1+ \frac{3}{5} \epsilon) \cos \theta/N_uN_{u^c}
\end{array} \right),
\end{equation}
\begin{equation}
D = aTv'(I + \Delta_{d^c})^{-\frac{1}{2}}
\left( \begin{array}{ccc}
0 & -c_{12} & -c_{13}/N_d \\
c_{12} & 0 & (\frac{1}{5} \epsilon \sin \theta - c_{23})/N_d \\
c_{13}/N_{d^c} & ((1+ \frac{2}{5} \epsilon) \sin \theta +c_{23})/N_{d^c} &
(1 + \frac{3}{5} \epsilon) \cos \theta/N_d N_{d^c}
\end{array} \right),
\end{equation}
\begin{equation}
L = aTv' \left( \begin{array}{ccc}
0 & c_{12} & c_{13}/N_{l^-} \\
-c_{12} & 0 & (-\frac{3}{5} \epsilon \sin \theta + c_{23})/N_{l^-} \\
-c_{13}/N_{l^+} & ((1+ \frac{6}{5} \epsilon) \sin \theta - c_{23})/N_{l^+} &
(1 + \frac{3}{5} \epsilon) \cos \theta/ N_{l^+}N_{l^-}
\end{array} \right) (I + \Delta_{l^-}^T)^{-\frac{1}{2}},
\end{equation}
\begin{equation}
N = N_0 = aTv \left( \begin{array}{ccc}
0 & 0 & 0 \\
0 & 0 & -\frac{3}{5} \epsilon \sin \theta/N_{\nu} \\
0 & -\sin \theta/N_{\nu^c} & -(1+ \frac{3}{5} \epsilon)
\cos \theta/N_{\nu^c}N_{\nu}
\end{array} \right).
\end{equation}

One sees that $U_0$ and $N_0$ are unaffected by adding $W_{vector}$
and that the effect on $D_0$ and $L_0$ is to add to them an antisymmetric
piece, $c_{ij}$, and to multiply them by mixing matrices
$(I + \Delta)^{-\frac{1}{2}}$. For the moment we will assume that
these factors can be neglected as would be the case if $\Delta_{d^c},
\Delta_{l^-} \ll 1$, although such matrices will play an important role
in our later discussion.  (Their exact forms are given by
$(\Delta_{d^c})_{ij} = \left|{{\langle {\bf 1(16)}_H} \rangle \over
d \langle {\bf 45}^{(X)}_H
\rangle }\right| ^2 p_i p_j (c_i^* c_j + c_i^{\prime *} c_j^\prime)$
where $\vec{p}
\equiv (1,1,N_{d^c}^{-1})$ and the same for $\Delta_{l^-}$ with $N_{d^c}$
replaced by $N_{l^-}$.)
The origin of the antisymmetric pieces, $c_{ij}$, can be
understood from Fig. 2.

\begin{picture}(360,216)
\thicklines
\put(36,144){\vector(1,0){36}}
\put(72,144){\line(1,0){72}}
\put(180,144){\vector(-1,0){36}}
\put(180,144){\vector(1,0){36}}
\put(216,144){\line(1,0){72}}
\put(324,144){\vector(-1,0){36}}
\put(108,96){\line(0,1){12}}
\put(108,114){\vector(0,1){12}}
\put(108,132){\line(0,1){12}}
\put(252,96){\line(0,1){12}}
\put(252,114){\vector(0,1){12}}
\put(252,132){\line(0,1){12}}
\put(180,96){\line(0,1){12}}
\put(180,126){\vector(0,-1){12}}
\put(180,132){\line(0,1){12}}
\put(175.5,94){$\times$}
\put(103.5,94){$\times$}
\put(247.5,94){$\times$}
\put(94.5,78){$\langle {\bf 16}_H \rangle$}
\put(238.5,78){$\langle {\bf 16}_H \rangle$}
\put(166.5,78){$\langle {\bf 45}_H^{(X)} \rangle$}
\put(63,153){${\bf 16}_i$}
\put(103.5,153){$c_i$}
\put(139.5,153){${\bf 10}$}
\put(175.5,153){$d$}
\put(207,153){${\bf 10}'$}
\put(247.5,153){$c_j'$}
\put(279,153){${\bf 16}_j$}
\put(162,42){{\bf Fig. 2}}
\end{picture}

\noindent
By inspection of that diagram one sees that
$W_{vector}$ contributes
\begin{equation}
\Delta W_{mass} = \sum_{i,j=1}^3(c_i c_j^{\prime} - c_i^{\prime} c_j)
\frac{\langle {\bf 1}({\bf 16}_H) \rangle \langle {\bf \overline{5}}
({\bf 16}_H)\rangle }
{d \langle {\bf 45}_H^{(X)} \rangle } {\bf 10}({\bf 16}_i) {\bf \overline{5}}
({\bf 16}_j).
\end{equation}
The antisymmetry in flavor is due to the antisymmetry of the adjoint
of $SO(10)$. Under the interchange of ${\bf 10}$ and ${\bf 10}'$ in
Fig.2 the diagram changes sign and $ij \rightarrow ji$. Comparing with
eqs.(15) and (16) one obtains $c_{ij} = (c_ic_j' - c_i'c_j)
\frac{M \langle {\bf 1}({\bf 16}_H)\rangle }{abd \langle {\bf 45}_H
\rangle \langle {\bf 45}_H^{(X)}\rangle }
\frac{\langle {\bf \overline{5}}({\bf 16}_H)\rangle }{\langle
{\bf \overline {5}}
({\bf 10}_H)\rangle }$.
We assume that the vacuum expectation value of the
${\bf 45}_H^{(X)}$ is in the $X$ direction so that the antisymmetric
pieces added to $D$ and $L$ are the same.
The smallness of the $c_{ij}$ can come from the smallness of
the $c_i$ or $c_i'$, or of their cross product, or of the ratios of
vacuum expectation values.

The forms of the full matrices exhibited in eqs.(14) -- (17) now
explain the several further relationships {\bf (V)}, {\bf (VI)}, and
{\bf (VIII)}.  Relation {\bf (V)}, that $|m_e^0/m_d^0|
\cong |m_s^0 m_b^0/m_{\mu}^0 m_{\tau}^0| \cong 1/3$, is a consequence
of $\det D = \det L$, which, it should be noted, is an {\it exact}
relation for
any values of the parameters both in the limit $\Delta \rightarrow
0$ and in the limit $N_f \rightarrow 1$ (which makes $\Delta_{d^c}
= \Delta_{l^-}$).

Relation {\bf (VI)}, that $m_u^0$ is proportionately
very tiny compared to $m_d^0$ and $m_e^0$, comes from the fact
that $W_{vector}$ contributes only to $D$ and $L$ and leaves $U$
rank 2, and thus $u$ massless.
Some higher order effects presumably give a small mass
to $u$, but as these contributions to $U$ would typically be of
order $10^{-5} aTv$ one would expect them to have
a negligible effect on everything else besides $m_u$. (For example,
their contribution to the Cabibbo angle would typically be of order
$m_u/m_c \sim 0.005$.)

Relation {\bf (VIII)}, $\tan \theta_c \cong \surd
\overline{m_d^0/m_s^0}$ is famously successful.$^5$ In schemes of the
Fritzsch type$^{13}$ there is an extra contribution $e^{i \alpha}
\surd \overline{m_u^0/m_c^0}$, which, being of magnitude
$0.07 \cong \frac{1}{3} \tan \theta_c$ requires that the phase
$\alpha$ take a particular value to obtain numerical agreement.
Here, because $W_{vector}$ does not contribute to $U$, one avoids
that extra term.

The foregoing three relations are consequences of the facts that $W_{vector}$
only contributes to $D$ and $L$ (required for {\bf (VI)} and {\bf (VIII)})
and that its contribution is antisymmetric (required for {\bf (V)} and
{\bf (VIII)}).
But it should be emphasized that these facts in turn are consequences
of aspects of $SO(10)$ group theory: namely, that the vector representation
contains only down quarks and leptons, and that the adjoint representation
is antisymmetric.

The final relation, {\bf (IX)}, $V_{ub} \sim V_{cb} V_{us}$,
follows if one makes the natural assumption that all the
$c_{ij}$ are of the same order,
since $V_{ub} \sim c_{13}/m_b^0$, $V_{us} \sim c_{12}/m_s^0$,
and $V_{cb} \sim m_s^0/m_b^0$. This assumption also would
imply that $c_{23}$ plays a negligible role.
So far, then, there are seven relevant combinations of
parameters$^{14}$: $aTv$, $v/v'$, $\sin \theta$, $\epsilon$, $c_{12}$,
$c_{13}$, and $N$. (And for small $T$, $N \cong 1$ and plays no role.)

Now comes the question of why $(m_c^0/m_t^0)/(m_s^0/m_b^0)
\stackrel{_<}{_\sim} \frac{1}{5}$,
when according to naive $SO(10)$ and eqs.(3), (4), and (12) it
ought to be approximately equal to unity.
There are two possible simple answers, that $m_b^0$ or $m_c^0$
is suppressed, and they lead to the two versions of
the model referred to earlier.
The first version, where $m_b^0$ is suppressed, has greater economy
as it makes do with only the terms in eq.(1).
In this version the factor $(I+ \Delta_{d^c})^{-\frac{1}{2}}$
which multiplies $D_0$ on the left in eq.(15) is assumed to have
the approximate form diag(1, 1, $\delta$) where $\delta
\stackrel{_<}{_\sim} \frac{1}{5}$. Then $m_b^0
\cong (D_{32}^2 + D_{33}^2)^{\frac{1}{2}}$ gets multiplied by $\delta$,
while $m_s^0 \cong D_{23} D_{32}/D_{33}$ is left unaffected. While this
works, it turns out to have some minor drawbacks.
The angle $\theta$ ends up being somewhat small, and, furthermore,
for $(I+\Delta_{d^c})^{-\frac{1}{2}}$ to have the desired form the
vector $c_i$ (or alternatively $c_i^{\prime}$) must also be nearly aligned
with $b_i$. Such a preferred direction in family space is somewhat
unappealing since family symmetry has been eschewed. A second drawback
is that $m_b^0 \cong m_{\tau}^0$ is no longer automatic but must be fit.
Aside from this, this version of the model preserves the explanatory
successes {\bf (I)} -- {\bf (IX)} and has in addition several interesting
predictions.$^{15}$
\begin{equation}
\begin{array}{ccl}
{\rm tan}\beta & \cong & v/v'  \cong   m_c^0/m_s^0, \\
|\theta_{e \mu}| & \cong & \surd \overline{m_e^0/m_{\mu}^0}
+ \frac{1}{2} \tan \theta [Re(\theta_{e \tau})/(1+
|\theta_{e \tau}|^2)], \\
\sin^2 2 \theta_{\mu \tau} & \stackrel{_>}{_\sim} & \frac{1}{2}.
\end{array}
\end{equation}

\noindent
Note that since the ratio $m_t^0/m_b^0$ has been affected by the
factor $(I+ \Delta_{d^c})^{-\frac{1}{2}}$, $\tan \beta$ is
given by the (unaffected) ratio $m_c^0/m_s^0$, which is quite
a distinctive prediction. The largeness of $\theta_{\mu \tau}$
arises because $L_{32}/L_{33}$
is no longer $\cong \tan \theta$ but (a careful calculation shows)
$\cong \frac{1}{2} \delta^{-1} \tan \theta$,
whereas $N_{32}/N_{33} \sim \tan \theta$.

We shall now discuss the second and in some ways cleaner version of the
model, in which $m_c^0$ is suppressed.  We shall see
that if the terms $W_{adjoint}$ given in eq.(2) are added to the
superpotential then the ratio $m_c^0/m_t^0$ is typically suppressed
relative to $m_s^0/m_b^0$ by a factor of $O(\epsilon)$
{\it without any special alignment of vectors
in family space or special values of parameters}. In other words,
the smallness of $m_c^0/m_t^0$ has a group-theoretical origin.
This suppression is achieved by a matrix $(I+ \Delta_{u^c})
^{-\frac{1}{2}}$ appearing in eq.(14) analogous to the matrices
appearing in eqs.(15) and (16).

Since $\langle {\bf 1}(\overline{{\bf 16}}_H) \rangle \neq 0$
and is $O(M_{GUT})$, $W_{adjoint}$ causes the
${\bf 10}({\bf 16}_i)$ to mix with the ${\bf 10}({\bf 45})$
so:
$\overline{{\bf 10}}({\bf 45}') \left[
f \Omega Q_{{\bf 10}({\bf 45})} {\bf 10}({\bf 45}) +
\langle {\bf 1}
(\overline{{\bf 16}}_H) \rangle e_i' {\bf 10}({\bf 16}_i) \right] $,
with similar mixing of ${\bf 10}({\bf 16}_i)$ and ${\bf 10}({\bf 45}')$
caused by the $e_i$ term. In particular,
since the ${\bf 10}({\bf 16}_i)$ contains the $u^c_i$
there will be produced in eq.(14) a mixing matrix $(I+ \Delta_{u^c})
^{-\frac{1}{2}}$ multiplying $U_0$ on the left, with $\Delta_{u^c}$
given by
\begin{equation}
(\Delta_{u^c})_{ij} =
\left| \frac{\langle {\bf 1}(\overline {{\bf 16}}_H)
\rangle }
{f \Omega Q_{u^c({\bf 45})}}\right|
^2 \left[ E_i^* E_j + E_i^{\prime *}
E_j^{\prime} \right],
\end{equation}

\noindent
where $E_i \equiv p_i e_i$, $E_i' \equiv p_i e_i'$, and $p_i \equiv
(1, 1, N_{u^c}^{-1})$.
(Compare with the expression for $\Delta_{d^c}$ given earlier.)
Since the ${\bf 10}({\bf 16}_i)$ also contains the $u_i$,
$d_i$, and $l_i^+$, there will be analogous mixing matrices
$(I+ \Delta_u)^{-\frac{1}{2}}$, $(I+ \Delta_d)^{-\frac{1}{2}}$,
and $(I+ \Delta_{l^+})^{-\frac{1}{2}}$ introduced in the obvious places
in eqs.(14) -- (16), where $\Delta_u$, $\Delta_d$, and $\Delta_{l^+}$
are given by eq.(20) with
$Q_{u^c({\bf 45})}$ being replaced by $Q_{u({\bf 45})}$, $Q_{d({\bf 45})}$,
and $Q_{l^+({\bf 45})}$, respectively, and $N_{u^c}$ replaced by the
appropriate $N$'s.

What is required to suppress $m_c^0/m_t^0$ is that at least some
elements of $\Delta_{u^c}$ should be quite large, whereas to preserve
the successful relations {\bf (I)} -- {\bf (IX)} the elements
of $\Delta_u$, $\Delta_d$, and $\Delta_{l^+}$ should be somewhat
(but not necessarily very much) smaller than unity. The required
largeness of $\Delta_{u^c}$ is explained rather elegantly by group
theory. The crucial
point is that the $Q_{u^c({\bf 45})}$ appearing squared in the
denominator of eq.(20) is {\it not} the $Q$ charge given in eq.(12) of the
$u^c_i$ that is contained in the ${\bf 16}_i$. Rather, it is the $Q$
charge of the $u^c$ that are contained in the
${\bf 45}$ and ${\bf 45}'$.
The difference is that $X_{{\bf 10}({\bf 45})} = -4$, whereas $X_{{\bf 10}
({\bf 16})} =1$, so that by eq.(11)  $Q_{{\bf 10}({\bf 45})}
= Q_{{\bf 10}({\bf 16})} + 1$.
{}From eq.(12), then,
one sees that $Q_{u^c({\bf 45})} = -\frac{4}{5} \epsilon$, and
therefore $\Delta_{u^c} \propto (\frac{1}{\epsilon})^2$. The expressions
for $\Delta_u$, $\Delta_d$, and $\Delta_{l^+}$, on the other hand,
have in their denominators $Q_{u({\bf 45})}^2 = Q_{d({\bf 45})}^2 =
(1 + \frac{1}{5} \epsilon)^2$ and $Q_{l^+({\bf 45})}^2 = (2+ \frac{6}{5}
\epsilon)^2$ and consequently are not enhanced.

It might be thought that some special form of $\Delta_{u^c}$ might
have to be assumed to get a suppression of $m_c^0/m_t^0$.
Curiously, owing to the rank-2 nature of $\Delta_{u^c}$,
this is not so. Generically, with the elements of
$\Delta_{u^c} = O(\frac{1}{\epsilon^2})$ and given
by eq.(20), $(I+ \Delta_{u^c})^{-\frac{1}{2}}$ suppresses
$m_c^0/m_t^0$ by a factor $O(\epsilon)$. More precisely,
if $\left| \vec{E} \right| \sim \left| \vec{E}' \right| > \epsilon$,
then $(m_c^0/m_t^0)/(m_s^0/m_b^0)
\cong \frac{4}{5} \epsilon \left| \frac{f \Omega} {\langle
{\bf 1}(\overline{{\bf 16}}_H) \rangle} \right| \frac
{\left| \vec{E} \times \vec{E}' \right| [E_1^2 + E_1^{\prime 2}]
^{\frac{1}{2}}}
{[(\vec{E} \times \vec{E}')_3]^2} = O(\epsilon)$.

It is assumed that while $\langle {\bf 1}(\overline {{\bf 16}}_H)
\rangle$ is
non-zero and superlarge, $\langle {\bf 5}(\overline {{\bf 16}}_H)
\rangle$ vanishes.
This condition can be naturally achieved$^{16}$ and means that $W_{adjoint}$
will contribute no antisymmetric piece to $U_0$. The only effect, then,
of $W_{adjoint}$ is the suppression of $m_c^0/m_t^0$. All the
successful relations {\bf (I)} -- {\bf (IX)} remain essentially
untouched. In particular, the effect of $(I+ \Delta_{u^c})^{-\frac{1}{2}}$
on $V_{cb}^0$ is to make $V_{cb}^0 \cong \frac{2}{5} \epsilon \sin \theta
\cos \theta \left[ 1 + \frac{N}{2} \tan \theta \frac{(\vec{E} \times
\vec{E}')_2}{(\vec{E} \times \vec{E}')_3} \right]$.
This version of the model has the following
predictions.
\begin{equation}
\begin{array}{ccl}
{\rm tan}\beta & \cong & v/v'  \cong  m_t^0/m_b^0, \\
\theta_{e \mu} & \cong & \surd \overline{m_e^0/m_{\mu}^0} + O(m_u^0/m_c^0), \\
\theta_{e \tau} & \cong & V_{ub} + O(m_u^0/m_t^0). \\
\end{array}
\end{equation}

\noindent
$\theta_{\mu \tau}$ gets a large contribution from $L_{32}/L_{33}
\cong \tan \theta \sim 1$, and can therefore be very large.
But for a wide range of the parameters describing $M_R$, the
Majorana mass matrix of the right-handed neutrinos,
$\theta_{\mu \tau} \cong
3 V_{cb}$ (see Ref. (17)) due to the near cancellation between $L_{32}/L_{33}$
and $N_{32}/N_{33}$.
More precisely, for $(M_R^{-1})_{22} \stackrel{_<}{_\sim}
(M_R^{-1})_{23} \stackrel{_<}{_\sim} (M_R^{-1})_{33}$,
$\theta_{\mu \tau} \cong 3 V_{cb} (1+ \frac{1}{2} [(M_R^{-1})_{23}/
(M_R^{-1})_{33}] N \tan \theta)$.
It should be noted, finally, that in
this version of the model, $\theta$ comes out to be of order 1,
so that there is no unexplained alignment in family space,
and $\epsilon$ comes out of order $\frac{1}{10}$. (This can
be seen from the relation $m_s^0/m_b^0 \cong \frac{N}{5} \epsilon
\sin^2 \theta$ and the relation just given for $V_{cb}^0$.)

In conclusion, we have shown that the group theory of $SO(10)$ can
elegantly explain the pattern of fermion masses
and mixing angles without family symmetry.  We find it remarkable that
the nine features listed in the introduction can arise as a
consequence of the simple Yukawa terms of eq.(1).
We also find it remarkable that the single group-theoretical
assumption $Q \sim I_{3R}$ explains three of those relations
as well as the smallness of $m_c^0/m_t^0$. The fact that the gauge
group is $SO(10)$ has played several crucial roles in the model:
it relates the up quarks and neutrinos to the down quarks and leptons,
it allows the VEVs of the adjoint Higgs fields to point in the $I_{3R}$
and $X$ directions, and it makes possible the antisymmetric
contributions to the mass matrices coming from $W_{vector}$.
We have only discussed terms in the superpotential that
are directly relevant to understanding the pattern of light
fermion masses. Other terms will be present including the Higgs
sector$^{12}$ and additional small Yukawa couplings$^{18}$,
but our results are not sensitive to these. Details of the
numerical fits and certain technical points will be presented in a
longer paper.

\section*{References}
\begin{enumerate}
\item  S. Dimopoulos,
L. Hall and S. Raby, Phys. Rev. Lett. {\bf 68}, 752 (1992); H. Arason,
D. Castano, E. Pirad and P. Ramond, Phys. Rev. {\bf D 47}, 232 (1992);
for a comprehensive review see
S. Raby, OSU preprint OHSTPY-HEP-T-95-024, hep-ph/9501349.
\item None of the discussions in this paper depends on supersymmetry,
although it will be assumed.
\item A. Buras, J. Ellis, M.K. Gaillard and D.V. Nanopoulos, Nucl. Phys.
{\bf B135}, 66 (1978).
\item H. Georgi and C. Jarlskog, Phys. Lett. {\bf 86B}, 297 (1979).
\item S. Weinberg, in {\it Festschrift for I.I. Rabi}, Trans. N.Y. Acad
. Sci. Ser. II {\bf 38}, 185 (1977); F. Wilczek and A. Zee, Phys. Lett.
{\bf 70B}, 418 (1977); H. Fritzsch, $ibid$., {\bf 70B}, 436 (1977);
R. Gatto, G. Sartori and M. Tonin, $ibid$., {\bf
28B}, 128 (1968); R.J. Oakes, $ibid.$, {\bf 29B}, 683 (1969).
\item This form can be enforced by a discrete $Z_3$ symmetry.
\item S.M. Barr, Phys. Rev. {\bf D 24}, 1895 (1981); B.S. Balakrishna,
A.L. Kagan and R.N. Mohapatra, Phys. Lett. {\bf B205}, 345 (1988);
Z. Berezhiani and R. Rattazzi, Nucl. Phys. {\bf B 407},
249 (1993);
K.S. Babu and E. Ma, Mod. Phys. Let. {\bf A4}, 1975 (1989) and
references therein.
\item S.M. Barr, Phys. Rev. Lett. {\bf 64}, 353 (1990).
\item More precisely $Q$ is in general a complex linear combination of
generators of $SO(10)$.
\item K.S. Babu and S.M. Barr, BA-94-45, hep-ph/9409285.
\item M. Srednicki, Nucl. Phys. {\bf B202}, 327 (1982).
\item S. Dimopoulos and F. Wilczek, Report No. NSF-ITP-82-07, August
1981 (unpublished); K.S. Babu and S.M. Barr, Phys. Rev. {\bf D48}, 5354
(1993); Phys. Rev. {\bf D50}, 3529 (1994).
\item H. Fritzsch, Nucl. Phys.
{\bf B155}, 189 (1979).
\item All parameters (eg. $ \epsilon, c_i, c_i^\prime, e_i, e_i^\prime$)
are assumed to be complex. With their phases of order unity sufficient
KM CP violation results.
\item tan$\beta \cong v/v^\prime$ holds only if the mixing of
$\overline{\bf 5}({\bf 16}_H)$ with the
$\overline{\bf 5}({\bf 10}_H)$ is small. The smallness of this mixing
can naturally explain why the $c_{ij} \ll 1$.
For a discussion of
gauge hierarchy including such mixing, see Ref. (16).
\item K.S. Babu and R.N. Mohapatra, Preprint BA-94-56, UMD-PP-95-57,
hep-ph/9410326.
\item A similar relation for $\theta_{\mu \tau}$ is derived in
K.S. Babu and R.N. Mohapatra, Phys. Rev. Lett {\bf 70}, 2845
(1993); but the prediction for $\theta_{e \tau}$ is different.
\item For example, small ${\bf 45}^2$ and ${\bf 45^\prime}^2$ mass
terms may be added which will make all components of ${\bf 45},
{\bf 45}^\prime$ superheavy and generate
large $\nu_i^c$ Majorana masses.
\end{enumerate}

\end{document}